\documentclass[aps,prl,twocolumn,showpacs,amsmath,prepintnumbers,amssymb,superscriptaddress,reprint]{revtex4}
\usepackage[dvipdfmx]{graphicx}
\usepackage{dcolumn}
\usepackage{color}
\usepackage[latin1]{inputenc}
\usepackage{comment}
\usepackage{amsmath}

\usepackage[normalem]{ulem}

\begin{document}
\preprint{Version v8}
%
\title{Direct Mapping of Spin and Orbital Entangled Wavefucntion under Interband Spin-Orbit coupling of Rashba Spin-Split Surface States
}
\author{Ryo~Noguchi}
\author{Kenta~Kuroda}
\thanks{kuroken224@issp.u-tokyo.ac.jp}
\author{K.~Yaji}
\affiliation{Institute for Solid State Physics, University of Tokyo, Kashiwa, Chiba 277-8581, Japan}
\author{K.~Kobayashi}
\affiliation{Department of Physics, Ochanomizu University, Bunkyo-ku, Tokyo 112-8610, Japan}
\author{M.~Sakano}
\author{A.~Harasawa}
\author{Takeshi~Kondo}
\author{F.~Komori}
\author{S.~Shin}
\affiliation{Institute for Solid State Physics, University of Tokyo, Kashiwa, Chiba 277-8581, Japan}
\date{\today}

%
\begin{abstract}                                                   %
%
We use spin- and angle-resolved photoemission spectroscopy (SARPES) combined with polarization-variable laser and investigate the spin-orbit coupling effect under interband hybridization of Rashba spin-split states for the surface alloys Bi/Ag(111) and Bi/Cu(111).
In addition to the conventional band mapping of photoemission for Rashba spin-splitting, the different orbital and spin parts of the surface wavefucntion are directly imaged into energy-momentum space. 
It is unambiguously revealed that the interband spin-orbit coupling modifies the spin and orbital character of the Rashba surface states leading to the enriched spin-orbital entanglement and the pronounced momentum dependence of the spin-polarization.
The hybridization thus strongly deviates the spin and orbital characters from the standard Rashba model.
The complex spin texture under interband spin-orbit hybridyzation proposed by first-principles calculation is experimentally unraveled by SARPES with a combination of $p$- and $s$-polarized light.
\end{abstract}
\maketitle
%
%
%
A realization of functional capabilities to generate spin-splitting of electronic states without any external magnetic field is a key subject in the research of spintronics~\cite{Manchon15NatureMat}.
A promising strategy exploits the influence of spin-orbit (SO) interaction that can give rise to the lifting of spin degeneracy under broken space inversion symmetry, the so-called Rashba effect~\cite{Rashba84JETP}.
In the conventional Rashba model, an eigenstate of the SO-induced spin-splitting is treated with an assumption of a pure spin state fully chiral spin-polarized which protects electrons from backscattering~\cite{Rashba84JETP,LaShell96prl, Nicolay01prb, Hoesch04prb}.
However, in real materials, the assumption can be usually broken because the SO coupling mixes different states with different orbitals and orthogonal spinors in a quasiparticle eigenstate~\cite{Henk03prb, Miyamoto16prb, Wissing14prl}. 
The SO entanglement can permit the spin-flip electron backscattering~\cite{Krasovskii15prl} and moreover orbital mixing in the eigenstate can play a significant role in an emergence of the large spin-splitting~\cite{Nagano09jpcm,Ishida14prb, Prak14jsrp,Bentmann09epl,Gierz10prb}.
Therefore, it is essentially important to experimentally explore the SO coupling not only in the lifting spin degeneracy but also in the spin and orbital wavefunction as eigenstates. 

\begin{figure}
\begin{center}
\includegraphics[width=0.94\columnwidth]{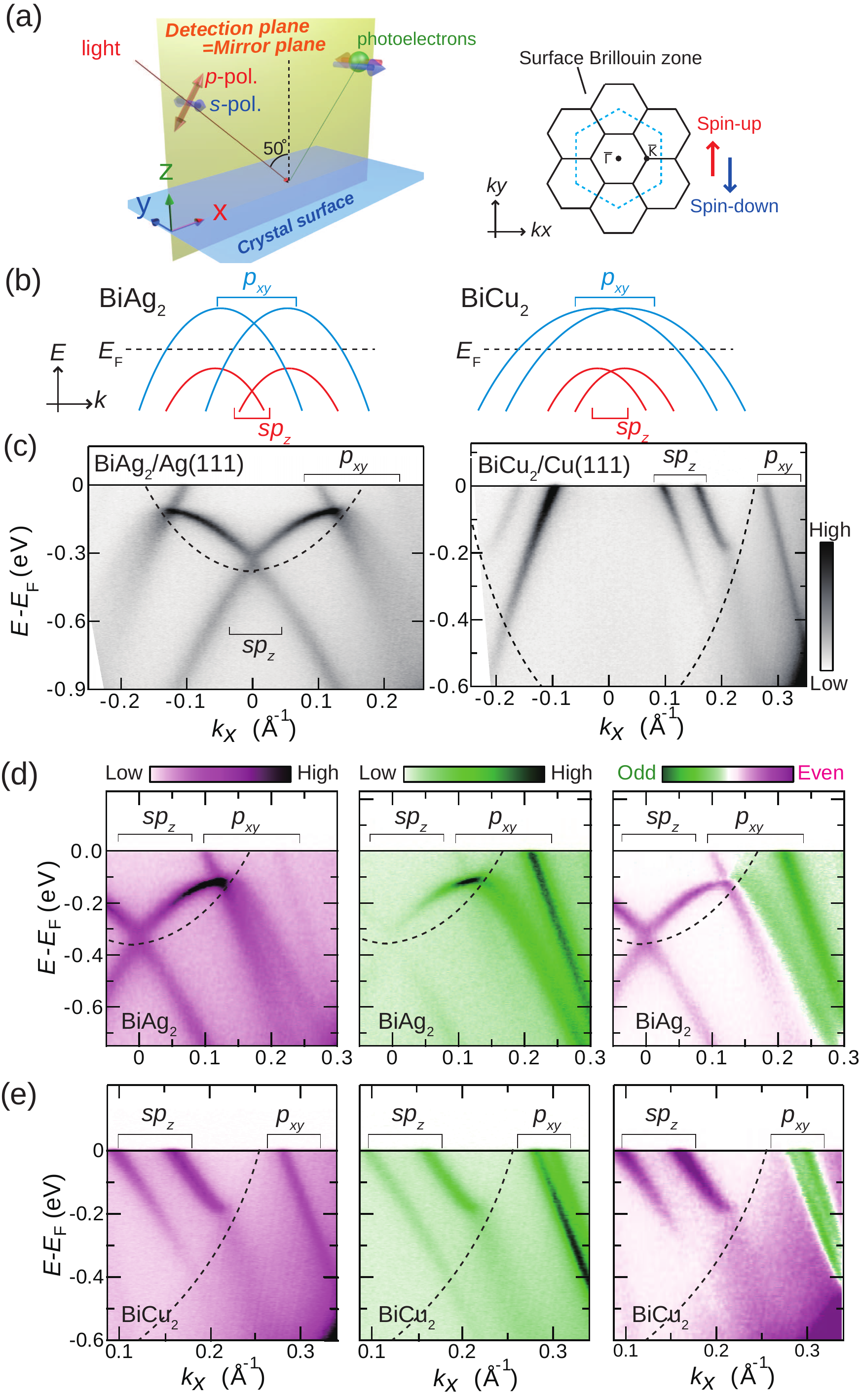}
\caption[]{(Color online) (a) The experimental configuration for $p$- or $s$-polarization with an angle incidence of 50~$^{\circ}$.
A mirror plane of the surface coincides with the plane of incidence ($x$-$z$ plane).
Surface Brillonin zone of the surface alloys [underlying fcc(111) substrate of Ag(111) and Cu(111)] is shown by solid (dashed) line.
(b) Schematic of the energy dispersion of Rashba spin-split bands in (left) BiAg$_{2}$ and (right) BiCu$_{2}$ surface alloys.
(c) ARPES intensity maps for BiAg$_{2}$ and BiCu$_{2}$ surface alloys with $p$-polarization along high symmetry $\bar{\rm{\Gamma}}$-$\bar{\rm{K}}$ line of the surface Brillouin zone.
The dashed lines indicate the edge of the projected bulk bands.
(d) and (e) The magnified ARPES intensity maps with (left) $p$- and (middle) $s$-polarization, and (right) the differential intensity maps, which are obtained by $I_{p}-I_{s}$ where $I_{p}$ and $I_{s}$ are the photoelectron intensity obtained by $p$- and $s$-polarized light without normalization, respectively.
}
\label{fig1}
\end{center}
\end{figure}

Beyond the conventional Rashba model, a well-ordered surface alloy BiAg$_{2}$ grown on Ag(111) provides an ideal case to study the SO entanglement in Rashba surface states.
In the surface alloy, an occupied $sp_{z}$-like band and a mostly unoccupied $p_{xy}$-like band show significant Rashba spin splitting~\cite{Ast07prl,He08prl} and cross each other at the specific $k_{||}$~\cite{Bihlmayer07prb,Bentmann12prl,Bian13prb} as shown in Fig.~\ref{fig1}~(b).
In particular, density functional theory (DFT) calculations showed the strong SO entanglement~\cite{Krasovskii15prl, Wissing14prl} and predicted the complex spin texture that is significantly different from the conventional Rashba model; the $sp_{z}$ band switches spin-polarization at the crossing through SO-induced interband hybridization~\cite{Bihlmayer07prb} which is in contrast to the similar system BiCu$_{2}$/Cu(111) as shown in Fig.~\ref{fig1}~(b)~\cite{Bentmann09epl,Mirhosseini09prb, Bentmann11prb}.
While the presence of the spin-polarized electronic bands has been demonstrated by spin and angle-resolved photoemission spectroscopy (SARPES)~\cite{Meier08prb, He10prl, Bentmann11prb} and inverse-SARPES~\cite{Wissing14prl}, the SO entanglement and particularly the spin texture due to the interband SO coupling is still under discussion, because of the lack of the orbital selectivity in the previous experiments.
Up to now, the complex spin texture of these surface states was only indirectly detected by quantum interference mapping through scanning tunneling spectroscopy~\cite{Krasovskii15prl,Hirayama11prl,Bode13prl}.
Thus, no conclusive understanding of the phenomenon apart from the conventional Rashba model has been achieved yet.

In this Rapid Communication, we directly investigate the SO entanglement in the Rashba surface states of BiAg$_{2}$/Ag(111) surface alloy by using a combination of polarization variable laser with SARPES (laser-SARPES) and compare to those of BiCu$_{2}$/Cu(111) as a simple case.
In contrast to the previous experiments~\cite{Meier08prb, He10prl, Bentmann11prb,Wissing14prl}, our laser-SARPES deconvolves the orbital wavefucntion and the coupled spin, and the surface wavefunctions are directly imaged into momentum-space through orbital-selection rule. 
It is shown that the interband SO coupling modifies the spin and orbital character of the Rashba surface states leading to the spin-orbital entanglement and the $k_{||}$ dependence.
The resulting spin texture thus shows a large deviation from the conventional Rashba model.
The full spin information is experimentally unraveled only by a combination of $p$- and $s$-polarized light in accordance with a view of the SO entanglement.
%
%
%
%

The laser-SARPES measurement was performed at the Institute for Solid State Physics (ISSP), the Univesity of Tokyo, with a high-flux 6.994-eV laser and ScientaOmicron DA30L photoelectron analyzer~\cite{Yaji16rsi}.
The experimental configuration is shown in Fig.~\ref{fig1}(a).
The $p$- and $s$-polarizations ($\epsilon_p$ and $\epsilon_s$, respectively) were used in the experiment.
The photoelectrons were detected along $\bar{\rm{\Gamma}}$-$\bar{\rm{K}}$ line of the surface Brillouin zone.  
The spectrometer resolved the spin component along $y$, which is perpendicular to the mirror plane of the surface.
The sample temperature was kept at $\sim$15~K.
The instrumental energy (angular) resolutions of the setup is 2~meV (0.3$^{\circ}$) and 20~meV (0.7$^{\circ}$) for spin-integrated ARPES and SARPES, respectively.
The BiAg$_{2}$ and BiCu$_{2}$ surface alloys were obtained by the procedures presented in the literatures~\cite{Ast07prl,Bentmann09epl}.
Low-energy electron diffraction measurements confirmed the ($\sqrt{3}\times\sqrt{3}$)$R$30$^{\circ}$ reconstruction of the surface alloys.

First-principles calculations were performed using the VASP code~\cite{Kresse96prb}.
The projector augmented wave method~\cite{Blochl94prb} is used in the plane-wave calculation.
The generalized gradient approximation by Perdew, Burke, and Ernzerhof~\cite{Perdew96prl} is used for the exchange-correlation potential.
The spin-orbit interaction is included.
The atom positions of BiAg$_{2}$ are optimized.
Those of BiCu$_{2}$ are taken from the experimental data of Ref.~\cite{Kaminski05ss}.

%
%
%
%
Let us start with showing a brief overview of observed electronic structure of BiAg$_{2}$ and BiCu$_{2}$ surface alloys in Fig.~\ref{fig1}(c).
The $sp_{z}$-derived bands and the higher-lying $p_{xy}$ bands disperse downwards in energy with a large Rashba spin-splitting in the both materials.
Compared with the surface bands in BiAg$_{2}$, the most part of the surface bands in BiCu$_{2}$ is above the Fermi level ($E_F$).
These results are in good agreement with previous works~\cite{Ast07prl,He08prl,Bentmann12prl,Bentmann09epl}.
%
\begin{figure*}[t]
\begin{center}
\includegraphics[width=0.90\textwidth]{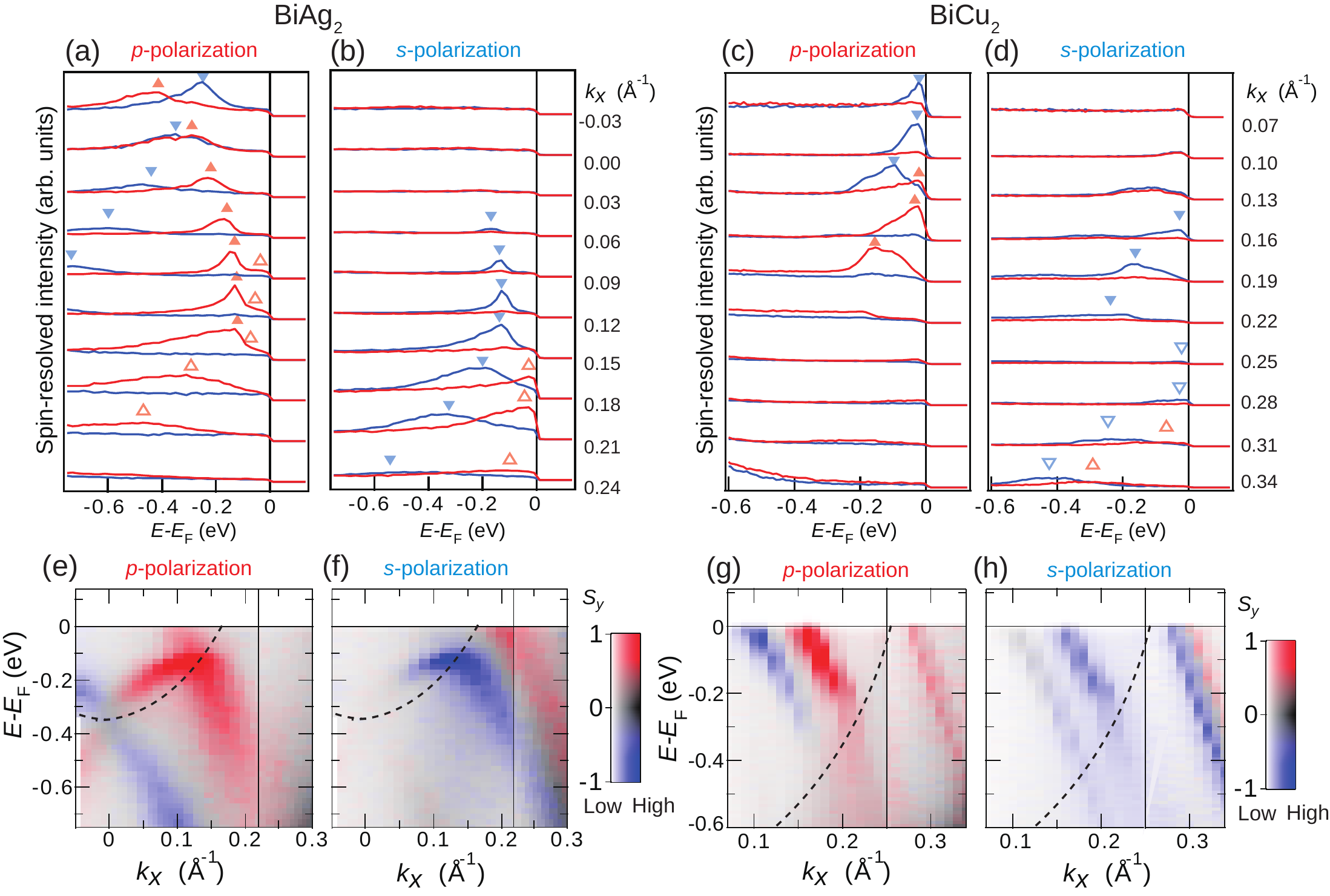}
\caption[]{(Color online) (a)-(d) SARPES spectra with spin quantum axis along $y$ for BiAg$_{2}$ and BiCu$_{2}$ by using $\epsilon_{p}$ and $\epsilon_{s}$.
Spin-up and spin-down spectra are plotted with red and blue lines.
The peak positions in the spin-resolved spectra for $sp_{z}$ and $p_{xy}$ bands are indicated by open and closed triangles, respectively.
(e)-(h) The corresponding spin-polarization and intensity maps with the two-dimensional color codes~\cite{Tushe15Ultra}.
The dashed lines indicate the edge of
the projected bulk bands.
}
\label{fig2}
\end{center}
\end{figure*}

Considering the orbital selection rule in the dipole excitation~\cite{Damascelli03rmp}, $\epsilon_p$ and $\epsilon_s$ enable us to draw different orbital symmetry.
Since the surface states near $E_{\rm{F}}$ are composed of Bi 6$s$ and 6$p$ orbitals~\cite{Bihlmayer07prb}, $\epsilon_p$ selectively detects weight of $even$-parity orbital with respect to the mirror plane, mainly from $s$, $p_{z}$ and $p_{x}$ components, while $\epsilon_s$ is sensitive to the $odd$-parity orbital mainly derived from $p_{y}$.

Figure~\ref{fig1}(d) summarizes the linear polarization dependence in BiAg$_{2}$.
For the result obtained by $\epsilon_p$ [see the left panel], we observe the strong intensity for the $sp_{z}$ and inner $p_{xy}$ bands.
The data particularly displays the band crossing of the outer $sp_{z}$ and inner $p_{xy}$ bands around $k_{||}$=0.15~$\rm{\AA}^{-1}$ [Fig.~\ref{fig1}(d)], where the spectral intensity of the outer $sp_{z}$ is strongly suppressed. 
Surprisingly, switching the light polarization $\epsilon_p$ to $\epsilon_s$, the spectral intensity is dramatically changed [see the middle panel in Fig.~\ref{fig1}(d)].
The dispersion of the outer $sp_{z}$ is clearly observed even at lager $k_{||}$$>$0.15$~\rm{\AA}^{-1}$ together with the outer $p_{xy}$ band.
Consequently, the overall parabolic dispersion of the outer $sp_{z}$ band is clearly seen, which was absent in previous experiments~\cite{Ast07prl,Bentmann09epl,Meier08prb}.
The right panel of Fig.~\ref{fig1}(d) shows the differential intensity map [see figure caption of Fig.1].
The red-blue color contrast reflects the contribution of the even- and odd-orbital components in the surface wavefunction.
It is immediately found that the orbital character of $sp_{z}$ band changes the orbital character at the band crossing. 

In contrast, the result for BiCu$_{2}$ is found to be simple [see Fig.~\ref{fig1}(e)]: the $sp_{z}$ and $p_{xy}$ bands comprise mainly even- and odd-parity orbitals, respectively.
These two bands in BiCu$_{2}$ are separated in momentum-space away from the band crossing.
Nevertheless, we see the reduction of the spectral intensity of the outer $sp_{z}$ band when it overlaps with the projected bulk-band, as observed also in the even-part of the outer $sp_{z}$ band in BiAg$_{2}$ [see the left panel in Fig.~\ref{fig1}(d)]. 
This common feature suggests the interaction of the the outer $sp_{z}$ band with the bulk $sp_{z}$ projection from the substrate~\cite{Bentmann09epl, He08prl} modifies the spectral weight of the orbital wavefunction particularly for the even-orbitals.

Apparently, the significant $k_{||}$ dependence of the orbital symmetry is unique for the outer $sp_{z}$ of BiAg$_{2}$.
This result indicates the presence of the interband SO hybridization that allows to mix the even- and odd-orbital components in the surface wavefunction of the outer $sp_{z}$ band.
Previously, it has been believed that the hybridization associates with gap opening~\cite{Bihlmayer07prb,Bian13prb,Bode13prl,Bentmann12prl}.
However, in our laser-SARPES experiment, there is no clear gap observed around the crossing point.
\begin{figure}[t!]
\includegraphics[width=0.99\columnwidth]{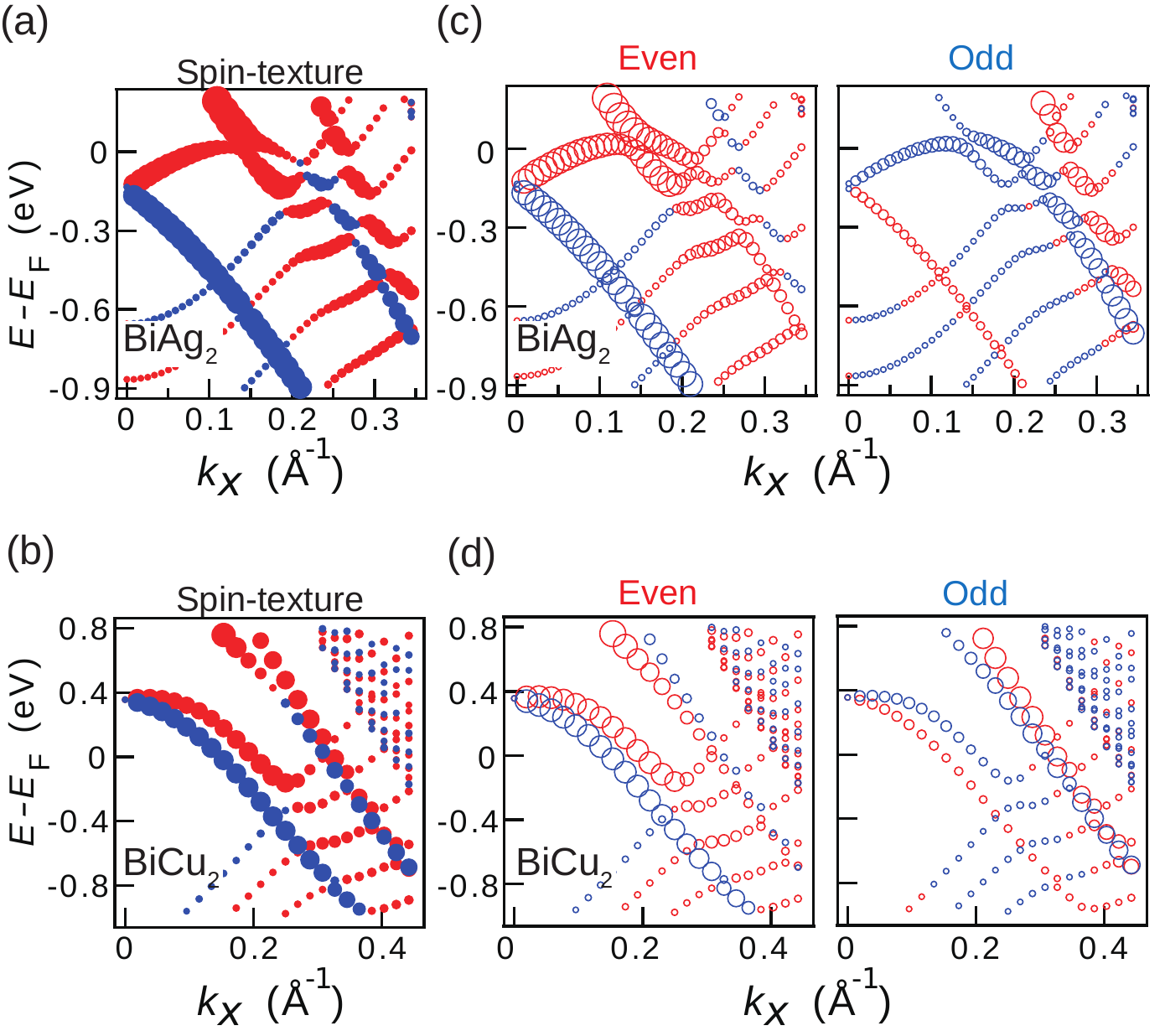}
\caption[]{(Color online)
(Color online) (a) and (b) Calculated total spin texture for BiAg$_{2}$ and BiCu$_{2}$ surface alloys, respectively.
(c) and (d) The coupled spin textures decomposed into (left) $even$- and (right) $odd$-orbital characters.
The size of the circles is proportional to the total spin polarizations.
The red and blue color indicate spin-up and spin-down quantized along $y$, respectively.
}
\label{fig3}
\end{figure}

To get further insight into the influence of the interband SO coupling, we carried out laser-SARPES measurements collaborated with the orbital selection rule of $\epsilon_p$ and $\epsilon_s$.
Figures~\ref{fig2}(a)-(d) show SARPES spectra obtained by using $\epsilon_{p}$ and $\epsilon_{s}$ for BiAg$_{2}$ and BiCu$_{2}$.
The corresponding spin-polarization and intensity maps~\cite{Tushe15Ultra} are shown in Figs.~\ref{fig2}(e)-(h).
For $\epsilon_p$, the inner and outer $sp_{z}$-bands around $\bar{\rm{\Gamma}}$ point in the both materials show negative and positive spin-polarization, respectively, displaying a conventional Rashba-type spin-texture~\cite{Meier08prb, He10prl}.
The observed spin polarization is found to be large up to nearly 80 and 60~$\%$ for BiAg$_{2}$ and BiCu$_{2}$, respectively.

Most remarkably, we find that the sign of the spin polarization sensitively depends on the linear polarization~\cite{foot}.
This can be seen in the SARPES spectra particularly for $k_{x}$=0.12~$\rm{\AA}^{-1}$ in BiAg$_{2}$.
The spectral weight of the spin-up is considerably larger than the spin-down for $\epsilon_{p}$ and achieves nearly $+$80~$\%$ spin-polarization.
For $\epsilon_{s}$, the intensity relation turns to the opposite and the resulting spin-polarization is found to be $-$70~$\%$.
Since the linear polarization is sensitive to different orbital symmetry, our laser-SARPES umambiguously reveals that the spin direction strongly depends on the orbital character.

Figures~\ref{fig3}(a) and (b) represent the calculated spin texture for the both materials, which is consistent with previous theoretical results~\cite{Bihlmayer07prb, Bentmann09epl,Mirhosseini09prb}.
In BiAg$_{2}$, the hybridization of the $sp_z$ and $p_{xy}$ bands is found in a gap opening at the band crossing where the spin polarization changes its sign [Fig.~\ref{fig3}(a)].
Due to the hybridization, one may not assign whether the two branches at the gapped point originate from either the $sp_{z}$ or the $p_{xy}$ derived states.
Nevertheless let us refer the $sp_{z}$ and $p_{xy}$ bands, since our experimental result shows the hybridization avoids the gap opening.

The complex spin textures are decomposed into the even and odd orbital contributions in Figs.~\ref{fig3}(c) and (d).
These results clearly show the SO entanglement in which the different orbital components are coupled with opposite spin.
The calculated SO entangled texture reproduces our experimental results for the spin mapping of the surface wavefunction (see Fig.~\ref{fig2}).
By general group-theoretical analysis for the mirror symmetry~\cite{Henk03prb}, the wavefucntion under SO coupling is generally represented as:
\begin{equation}
|\Psi_{\pm{i}}>=|even, \uparrow (\downarrow)>+|odd, \downarrow (\uparrow)>,
\label{group}
\end{equation} 
where the spinors $|\uparrow>$, $|\downarrow>$ are quantized along $y$, which is perpendicular to the mirror plane,  and the index $\pm{i}$ is representation for the mirror symmetry.
This explains not only that the $even$- and $odd$-parity orbitals couple with opposite spins but also that the SO entanglement is a general consequence of the SO coupling.
Indeed the similar SO-coupled states have been recently confirmed in surface states of topological insulators~\cite{Zhang13prl,Zhu13prl,Cao13NaturePhys,Zhu14prl, Xie14NatureComm, Kuroda16prb} and Rashba states in BiTeI~\cite{Bawden15sa,Maass16naturecom}.

In BiAg$_{2}$ [Figs.~\ref{fig3}(a) and (c)], the spin-polarization coupled to the even-orbital component predominates the $sp_{z}$ state around the $\bar{\rm{\Gamma}}$ point, and finally the opposite spin coupled to the odd-orbital component becomes dominant at higher $k_{||}$.
This indicates that the weight of the even- and odd-orbital components in the surface wavefunction play a significant role in the total spin texture through the SO entanglement [Fig.~\ref{fig3}(a)].
Our experiment indeed demonstrates the significant $k_{||}$-dependence of the orbital wavefunction [Fig.~\ref{fig1}(d)], which shows a good agreement with DFT calculation~\cite{si}.
Therefore, the hybridization through the interband SO coupling modifies the orbital component and induces the SO entanglement in the $sp_{z}$, which considerably deviates the spin texture from the conventional Rashba model.
\begin{figure}[t]
\begin{center}
\includegraphics[width=0.99\columnwidth]{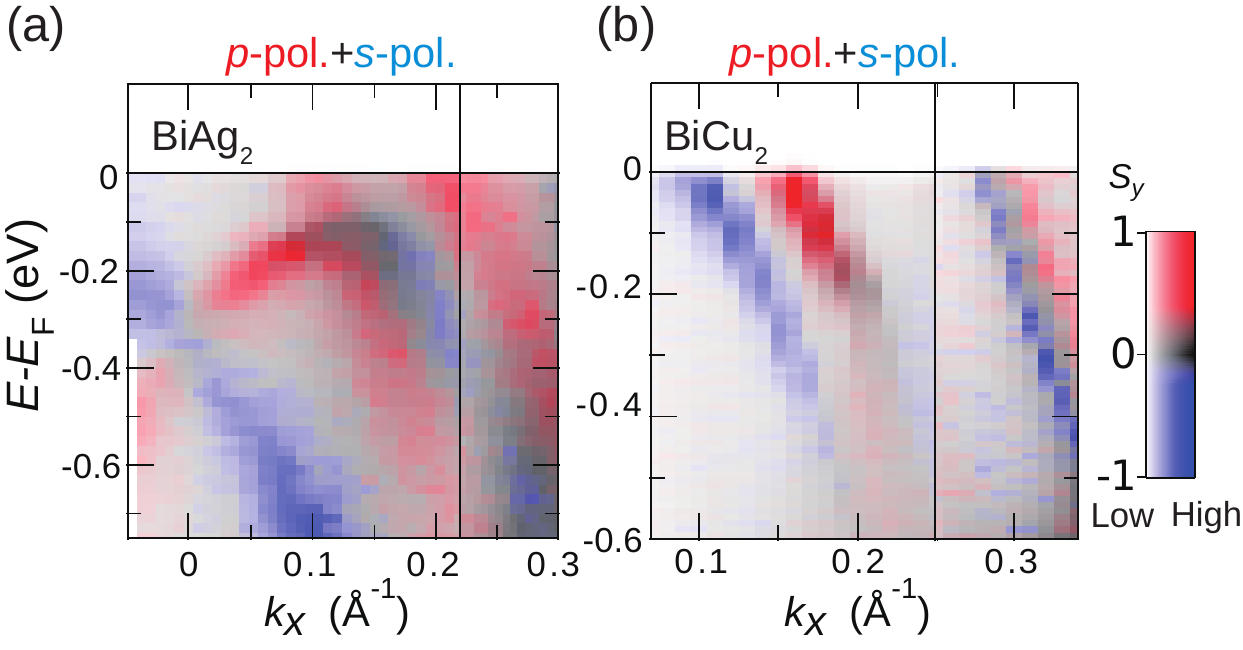}
\caption[]{(Color online)
(a) and (b) Experimentally obtained orbital-integrated spin-texture of the Rashba surface states in BiAg$_2$ and BiCu$_2$, which is in good agreement with that theoretically predicted as shown in Figures~\ref{fig3}(a) and (b).
The color arrangement indicates the total photoelecton intensity and spin-polarization~\cite{Tushe15Ultra}.
}
\label{fig4}
\end{center}
\end{figure}

Apparently, the mapping of the spin in our experiment [Fig.~\ref{fig2}(e)-(h)] does not reflect the predicted spin texture [Figs.~\ref{fig3}(a) and (b)]. 
This is because photomiesion measurement by using linearly polarized light in our experimental set-up selects the specific orbital symmetry~\cite{Wissing14prl}.
Indeed, the orbital-dependent spin texture in theory [Figs.~\ref{fig3}(c) and (d)] shows good agreement with our laser-SARPES results for $\epsilon_{p}$ and $\epsilon_{s}$.

We now show that the total spin information can be traced back only by using a combination of the spin mapping with $\epsilon_{p}$ and $\epsilon_{s}$ lasers.
Owing to selective detection of the pure orbital-symmetry in our experimental set-up of Fig.~\ref{fig1}(b), the orbital-dependence in the spin polarization is eliminated by integrations of spin-polarization maps ($P_{total}$) in Figs.~\ref{fig4}(a) and (b) as follows:
\begin{equation}
P_{total}=\frac{(I_{\uparrow, p}+I_{\uparrow, s})-(I_{\downarrow, p}+I_{\downarrow, s})}{(I_{\uparrow, p}+I_{\uparrow, s})+(I_{\downarrow, p}+I_{\downarrow, s})},
\end{equation} 
where $I_{\uparrow, p}$ ($I_{\uparrow, s}$) and $I_{\downarrow, p}$ ($I_{\downarrow, s}$) indicates the spin-up and spin-down intensities obtained by $\epsilon_{p}$ ($\epsilon_{s}$).
The mapping of $P_{total}$ clearly demonstrates the complex spin texture of the $sp_{z}$ band under interband hybridization, which is obviously comparable to the theoretical predictions in Figs.~\ref{fig3}(a) and (b).  
Since the SO entanglement is generally expected in materials as long as the SO coupling plays a significant role,
this technique therefore demonstrates a general advantage to investigate the unconventional spin textures in strong SO-coupled states although the results could be infuenced by the cross-section for $\epsilon_{p}$ and $\epsilon_{s}$ and require the photoemission calculation to consider photon energy dependence~\cite{Heinzmann2012spin,Krasovskii15spin}.

The question remains as to why a hybridization gap is absent in our experiment while the calculation predicts the gap.
We attribute the absence of the gap to an interaction of the confined surface states with the electronic state in the substrate.
It was recently shown that the size of the hybridization gap sensitively depends on the thickness of Ag(111) quantum well and decreases with increasing substrate thickness~\cite{Bentmann12prl}.
Hence, one can expect the gap absence in the case of the bulk substrate with continuum electronic states.
The similar hybridization reconstructed by the bulk interaction is reported in image potential resonances~\cite{Winter11prl}.
This fact indicates that the presence/absence of the gap does no give a direct evidence of the interband SO-coupling but the pronounced reconstruction of the surface wavefunction directly displays.
In particular, the knowledge of the interband SO coupling is critically important for emergence of Dirac and Weyl fermions in semiconductors and semimetals~\cite{Hasan10rmp, Ando13jpsj, Wan11prb, Burkov11prl, Hasan15Science, Hasan16Naturecom}, related to non-trivial band topology.

%
%
In conclusion, we have deconvolved the spin and orbital wavefuntion of the Rashba spin-split surface states for the Bi-based surface alloys, and directly mapped these wavefucntions into momentum-space by combining orbital-selective laser-SARPES and first principles calculations.
The interband SO hybridization strongly influence the spin and orbital character in the surface wavefunction leading to the $k_{||}$ dependence of the SO entanglement.
The resulting spin texture thus considerably deviates from the conventional Rashba model.
Although the measured spin texture by using $\epsilon_p$ or $\epsilon_s$ does not give the full spin information in this case, the full spin-texture is experimentally unraveled by SARPES with a combination of the both linear polarization.
Our findings can be widely applied for clarifying the complex spin information in the SO-entangled surface states.

%
%
We gratefully acknowledge funding JSPS Grant-in-Aid for Scientific Research (B) through Projects No. 26287061 and for Young Scientists (B) through Projects No. 15K17675.
%

\bibliographystyle{prsty}

\begin{thebibliography}{10}
\bibitem{foot} The inner $sp_z$ band in BiCu$_2$ does not show the orbital-dependence in Figs.~\ref{fig2}(g) and (h), which is conflict with the theoretical result in Fig.~\ref{fig3}(d).
We expect that this may indicate an interaction with the bulk projection bands.
However, it is strongly required further theoretical supports to fully explain this behavior.
\end{thebibliography}

\end{document}